# Deep Learning for Forecasting Stock Returns in the Cross-Section

Masaya Abe[1] and Hideki Nakayama[2]

[1] Nomura Asset Management Co., Ltd., Tokyo, Japan
m-abe@nomura-am.co.jp
[2] The University of Tokyo, Tokyo, Japan
nakayama@nlab.ci.i.u-tokyo.ac.jp

**Abstract.** Many studies have been undertaken by using machine learning techniques, including neural networks, to predict stock returns. Recently, a method known as deep learning, which achieves high performance mainly in image recognition and speech recognition, has attracted attention in the machine learning field. This paper implements deep learning to predict one-month-ahead stock returns in the cross-section in the Japanese stock market and investigates the performance of the method. Our results show that deep neural networks generally outperform shallow neural networks, and the best networks also outperform representative machine learning models. These results indicate that deep learning shows promise as a skillful machine learning method to predict stock returns in the cross-section.

## 1 Introduction

Stock return predictability is one of the most important concerns for investors. In particular, many authors attempt to explain the cross-section of stock returns by using various factors, such as earnings–price ratio, company size and stock price momentum, and the efficacy of using such factors [1-3]. Conversely, the investors themselves must decide how to process and predict return, including selection and weighting of such factors.

One way to make investment decisions is to rely upon the use of machine learning models. This is a supervised learning approach that uses multiple factors to explain stock returns as input values and future stock returns as output values. Deep learning has attracted attention in recent years in the machine learning field because of its high performance in areas such as image recognition and speech recognition [4, 5]. Deep learning is a representation-learning method with multiple levels of representation. This method passes data through many simple but nonlinear modules. The data passes

---

The views expressed here are our own and do not necessarily reflect the views of Nomura Asset management. Any errors and inadequacies are our own.

through many more layers than it does in conventional three-layer neural networks. This enables a computer to build complex concepts out of simpler concepts [4, 5].

By inputting data of multiple factors and passing them through many layers, deep learning could extract useful features, increase representational power, enhance performance, and improve the prediction accuracy for future stock returns. Currently, there have been few applications of deep learning to report on stock return predictability. Positive results of such applications could certainly be said to expand the versatility of the deep learning technique across multiple fields.

In this paper, we use deep learning to predict one-month-ahead stock returns in the cross-section in the Japanese stock market. We calculate predictive stock returns (scores) from the information of the past five points of time for 25 factors (features) for MSCI Japan Index constituents. As a measure of the performance, we use rank correlation between the actual out-of-sample returns and their predicted scores, directional accuracy, and performance of a simple long–short portfolio strategy. We compare with conventional three-layer neural networks and support vector regression and random forests as representative machine learning techniques.

## 2 Related Works

Many studies on stock return predictability have been reported on neural networks [6, 7]. Most of those are forecasts of stock market returns; however, forecasts of individual stock returns using the neural networks dealt with in this paper have also been conducted. For example, Olson and Mossman [8] attempted to predict one-year-ahead stock returns for 2,352 Canadian companies using 61 accounting ratios as input values and reported that neural networks outperform traditional regression techniques. As an application to emerging market, Cao et al. [9] predicted stock returns in the Chinese stock market. They showed that neural networks outperform the linear model. Besides those, Kryzanowski et al. [10] found that neural networks correctly classify 72% of the positive/negative returns to predict one-year-ahead stock returns by using financial ratios and macroeconomic variables.

Studies on deep learning have been recently undertaken due to the heightened attention toward this technique. Krauss et al. [11] used three different machine learning models, deep neural networks, gradient-boosted trees and random forests to predict one-day-ahead stock returns for the S&P500 constituents. As a result, they showed that combining the predictions of those three as an equal-weighted ensemble outperforms each individual model. Among each model, random forests outperform deep neural networks and gradient-boosted trees. Conversely, they stated that careful hyper-parameter optimization may still yield advantageous results for the tuning-intensive deep neural networks. Outside the stock market, Dixon et al. [12] attempted to predict the direction of instrument movement for 5-min mid-prices for 43 CME listed commodity and FX futures. They showed 68% accuracy for the high ones. Moreover, in an application to a simple trading strategy, the best instrument has an annualized Sharpe Ratio of 3.29, indicating its high prediction ability. These studies were implemented for short investment horizons and do not use financial variables as input values. The present paper predicts one-month-ahead stock returns using multiple factors from both market and financial data as input values.

## 3 Data and Methodology

### 3.1 Dataset for MSCI Japan Universe

We prepare dataset for MSCI Japan Index constituents. The MSCI Japan Index comprises the large and mid-cap segments of the Japanese market. As of January 2017, the index is composed of 319 constituents and covers approximately 85% of the free float-adjusted market capitalization in Japan [13]. The index is also often used as a benchmark for overseas institutional investors investing in Japanese stocks. We use the 25 factors listed in Table 1. These are used relatively often in practice. In calculating these factors, we acquire necessary data from WorldScope, Thomson Reuters, I/B/E/S, EXSHARE, and MSCI. The actual financial data is acquired from WorldScope and Reuters Fundamentals (WorldScope priority). Taking into account the time when investors are actually available, we have a lag of four months. Forecast data is obtained from Thomson Reuters Estimates and I/B/E/S Estimates (Thomson Reuters priority). The data is used to calculate the factors from No. 2 to No. 8 and Nos. 16 and 17. Factors are calculated on a monthly basis (at the end of month) from December 1990 to November 2016. Note that factor calculation is not performed for Nos. 18 and 24. We directly use “Historical Beta” for No.18 and “Predicted Specific Risk” for No.24 from the MSCI Barra JPE4 model. Stock returns with dividends are acquired on a monthly basis (at the end of month).

**Table 1.** List of factors.

| No. | Factor | No. | Factor |
|---|---|---|---|
| 1 | Book-to-market ratio | 14 | Investment growth |
| 2 | Earnings-to-price ratio | 15 | Investment-to-assets ratio |
| 3 | Dividend yield | 16 | EPS Revision(1 month) |
| 4 | Sales-to-price ratio | 17 | EPS Revision(3 months) |
| 5 | Cash flow-to-price ratio | 18 | Market beta |
| 6 | Return on equity | 19 | Market value |
| 7 | Return on asset | 20 | Past stock return(1 month) |
| 8 | Return on invested capital | 21 | Past stock return(12 months) |
| 9 | Accruals | 22 | Volatility |
| 10 | Sales-to-total assets ratio | 23 | Skewness |
| 11 | Current ratio | 24 | Idiosyncratic volatility |
| 12 | Equity ratio | 25 | Trading turnover |
| 13 | Total asset growth | | |

### 3.2 Problem Definition

To define the problem as a regression problem. For example, for stock $i$ in MSCI Japan Index constituents at month $T$ (end of month), 25 factors listed in Table 1 are defined by $\mathbf{x}_{i,T} \in R^{25}$ and input values are defined by $\mathbf{v}_{i,T} = \{\mathbf{x}_{i,T}, \mathbf{x}_{i,T-3}, \mathbf{x}_{i,T-6}, \mathbf{x}_{i,T-9}, \mathbf{x}_{i,T-12}\} \in R^{125}$ using the past five points of time in three month intervals for 25 factors. The output value is defined by the next month's stock return, $r_{i,T+1} \in R$. As a more specific example, Fig. 1 shows the relationship between the input values and the output value for stock $i$ from one set of training data at December 2001 as $T$+1. The set consists of all stocks in MSCI Japan Index constituents at November 2001 ($T$). The input values are as follows: November 2001 ($T$), August 2001 ($T-3$), May 2001 ($T-6$), February 2001 ($T-9$), and November 2000 ($T-12$), as factors of past five points of time. The output value is the actual stock return at December 2001 ($T$+1). For data preprocessing, rescaling is performed so that each input value is maximally 1 (minimum $\approx$ 0) by ranking each input value in an ascending order by stock universe at each time point and then dividing by the maximum rank value. Similar rescaling is done for output values $r_{i,T+1}$, to convert to the cross-sectional stock returns (scores). Note that $\mathbf{v}_{i,T}$ and $r_{i,T+1}$ are assumed to be the values after data preprocessing.

This procedure is extended to using the latest $N$ months rather than the most recent set of training data (one training set). We use the mean squared error (MSE) as the loss function and define $\mathrm{MSE}_{T+1}$ when training the model at $T$+1 as follows:

$$\mathbf{MSE}_{T+1} = \frac{1}{K}\left\{\sum_{t=T-N+1}^{T}\sum_{i\in U_t}\left(r_{i,t+1} - f\left(\mathbf{v}_{i,t};\boldsymbol{\theta}_{T+1}\right)\right)^2\right\} \quad (1)$$

In (1), $K$ is the number of all training examples. $U_t$ is the MSCI Japan Index universe at $t$. $\boldsymbol{\theta}_{T+1}$ is the parameter calculated by solving (1) and makes the form of a function $f(.)$.

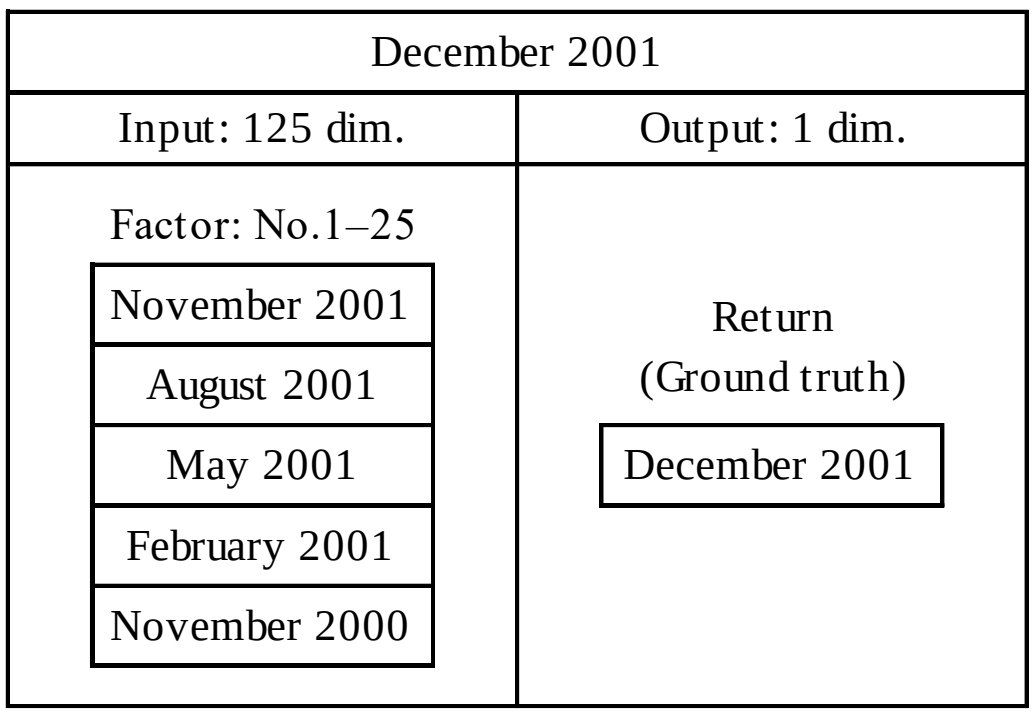

**Fig. 1.** Stock $i$ from one set of training data at December 2001.

### 3.3 Training and Prediction

We train the model by using the latest 120 sets of training from the past 10 years. To calculate the prediction, we substitute the latest input values into the model after training has occurred. The cross-sectional predictive stock return (score) of stock $i$ at time $T$+2 is calculated from time $T$+1 by (2) substituting $\mathbf{v}_{i,T+1}$ into the function $f(.)$ in (2) with the parameter $\boldsymbol{\theta}^*_{T+1}$, where $\boldsymbol{\theta}^*_{T+1}$ is calculated from (1) with $N = 120$:

$$Score_{i,T+2} = f\left(\mathbf{v}_{i,T+1}; \boldsymbol{\theta}^*_{T+1}\right) \tag{2}$$

For example, in order to calculate the prediction score at January 2002 ($T$+2) from December 2001 ($T$+1), the input values are as follows: December 2001 ($T$+1), September 2001 ($T$−2), January 2001 ($T$−5), March 2001 ($T$−8), December 2000 ($T$−11), as factors of the past five time points. The MSCI Japan Index constituents are from December 2001 ($T$+1). However, the prediction scores are not calculated for stocks with 63 or more missing input values, which is about half of the total number (125) of input values. For stocks with 62 or less missing input values, each missing value is replaced by the median value for the stocks that are not missing. For this series of processes, the model is updated by sliding one-month-ahead and carrying out a monthly forecast. The prediction period is 15 years: from January 2002 to December 2016 (180 months). An illustration of the flow of the processing is shown in Fig. 2, which shows the relationship between prediction and training data at each time point. For example, December 2001 in the “Training: 120 sets” is associated with Fig. 1 and January 2002 in the “Prediction: 1 set” represents the prediction for January 2002 from December 2001. The arrows indicate that the model is updated every month with the data sliding one-month-ahead.

### 3.4 Performance Measures

Rather than using the value of the loss function directly as a performance measure, we use the rank correlation coefficient (CORR) and directional accuracy (Direction) because these are more relevant measures of performance than the loss function. In addition, the performance of a simple long–short portfolio strategy is evaluated in comparison with support vector regression and random forests. In practice, these are used as methods to evaluate the performance of the cross-sectional stock returns. CORR is Spearman's rank correlation coefficient between the actual out-of-sample returns (next month's returns) and the prediction scores, which is used to measure the prediction accuracy of the entire predicted stock excluding the influence of outliers of individual stock returns.

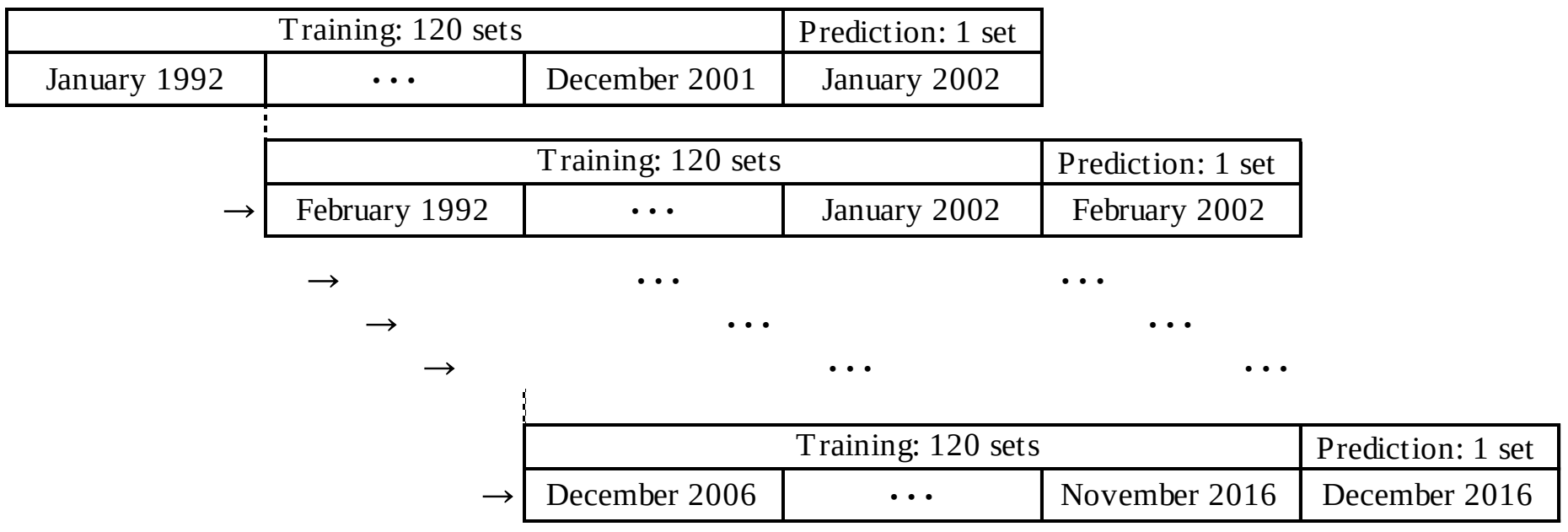

**Fig. 2.** Training-prediction set.

In an actual investment, there are many cases where the number of stocks is limited to those with higher prediction scores and those with lower prediction scores. We construct a portfolio comprising stock groups with top and bottom prediction scores. Direction is calculated by dividing the total number of the top stocks with high prediction scores that are above the cross-sectional median for next month's return on the stock universe and the bottom stocks with low prediction scores that are below the median by the total number of the top and bottom stocks.

The long–short portfolio strategy is a net zero investment strategy that buys the top stocks with equal weighting and sells the bottom stocks with equal weighting. To form into the top and bottom stock groups, we make two types of portfolios: tertile and quintile portfolios. These performance measures are calculated monthly during the prediction period. For example, at the evaluation starting point January 2002 (Prediction: 1 set in Fig. 2), these measures are calculated from the prediction scores for January 2002 from December 2001 and the actual out-of-sample returns at January 2002. Considering the stability of these evaluation results, it is necessary to consider a stock universe with at least a few dozen members in each category. Table 2 shows the monthly average numbers for the stock universe for the evaluation period from December 2001 to November 2016 with top and bottom stocks for both tertiles and quintiles. The total number of stocks exceeds 300; moreover, for quintiles, the total number of top and bottom stocks exceeds 100. Therefore, we consider that the size of the stock universe is adequate.

### 3.5 Compared Models

**Neural Networks.** All of the neural networks examined this paper are fully-connected feedforward neural networks. Table 3 shows all 16 types of the neural networks. The number in the "Hidden layers" column represents the number of units. For multiple layers, the layer numbers are connected with hyphens. We examine a total of

**Table 2.** Number of stocks (monthly average).

| **All** | **Tertile** | | **Quintile** | |
|---|---|---|---|---|
| | Top | Bottom | Top | Bottom |
| 336.5 | 112.5 | 111.8 | 67.7 | 66.9 |

8 patterns of deep neural networks (DNN) with 8 layers (DNN8) and with 5 layers (DNN5). The dropout rate is set uniformly to 50%. The number of units in each layer is designed to decrease as the layer becomes closer to the output layer. The patterns of DNN5 are designed to exclude duplicated layers of DNN8. For the conventional three-layer architectures, there are 8 patterns in total, 4 patterns with dropout rate set to 50% (NN3_DO) and 4 patterns with dropout rate set to 0% (NN3). For NN3_DO, the number of units of the hidden layer is adjusted so as to be approximately equal to the number of parameters (all weights including bias) of each pattern of DNN8. For example, the total number of parameters for NN3_DO_1, with 244 units in the hidden layer, is 30,989. This is approximately equal to 30,931 parameters for DNN8_1. For the number of units in the hidden layer of NN3, we select 4 large units in order from all the hidden layers of DNN8. As an intersection, we use the hyperbolic tangent as the activation function, Adam [14] for the optimization algorithm. The mini batch size is the size of the stock universe at each time point (approximately 300) with 100 epochs. We use TensorFlow for implementation. We initialize the biases to be 0 and generate the initial weight from TensorFlow's function "tf.truncated_normal" set to mean "0" and standard deviation "$1/\sqrt{M}$" (M is the size of the previous layer).

**Table 3.** Architectures of neural networks.

| **Neural Networks** | **Architectures** | | |
|---|---|---|---|
| | Number of layers | Hidden layers | Dropout |
| DNN8_1 | 8 | 100-100-50-50-10-10 | 50% |
| DNN8_2 | | 100-100-70-70-50-50 | |
| DNN8_3 | | 120-120-70-70-20-20 | |
| DNN8_4 | | 120-120-80-80-40-40 | |
| DNN5_1 | 5 | 100-50-10 | 50% |
| DNN5_2 | | 100-70-50 | |
| DNN5_3 | | 120-70-20 | |
| DNN5_4 | | 120-80-40 | |
| NN3_DO_1 | 3 | 244 | 50% |
| NN3_DO_2 | | 322 | |
| NN3_DO_3 | | 354 | |
| NN3_DO_4 | | 399 | |
| NN3_1 | 3 | 70 | 0% |
| NN3_2 | | 80 | |
| NN3_3 | | 100 | |
| NN3_4 | | 120 | |

**Support Vector Regression and Random Forests.** Support vector regression (SVR) and random forests (RF) are implemented with scikit-learn. The problem definition for SVR is ε-SVR [15] which is implemented with the class "sklearn.svm.SVR". For hyper-parameters C, gamma, epsilon, we implement 24 patterns of combinations of C = {0.1, 1.0, 10.0}, gamma = {0.0001, 0.001, 0.01, 0.1}, epsilon = {0.01, 0.1}. As an intersection, we use Radial Basis Function (RBF) as the type of kernel. We also define RF as a regression problem [16], and implement with the class "sklearn.ensemble.RandomForestRegressor" in scikit-learn. For hyper-parameters max_features, max_depth, we implement 37 patterns added by 16 patterns of combinations of max_features = {5, 10, 15, 20}, max_depth = {3, 5, 7, 9} and 21 patterns of combinations of max_features = {25, 30, 35}, max_depth = {3, 5, 7, 9, 11, 15, 20}. As an intersection, we set n_estimators (number of trees) to 1,000.

# 4 Experimental Results

## 4.1 Shallow versus Deep Neural Networks

Table 4 shows results from the neural networks patterns listed in Table 3. All values are monthly averaged. We have conducted a one-sided test of $H_0$: $p = 50\%$ against $H_1$: $p > 50\%$ for Direction. The best value for each set of 4 patterns is shown in bold, and the best value in each column is also underlined.

First, we look at CORR. DNN8_3 has the highest value of 0.0591, NN3_3 is the lowest at 0.0437, and values tend to increase as the number of layers increases. It can be confirmed that the DNN group outperforms even NN3_DO, which has had the number of units in its hidden layer adjusted to approximately match the number of parameters in DNN8.

The results for Direction are generally consistent with those for CORR and tend to be better as the number of layers increased in both the tertile and the quintile groups. Direction values reject the null hypothesis at the 0.1% significance level for all the patterns. Top and bottom quintiles had Direction values that are 0.4 to 1.0% higher than those for top and bottom tertiles.

The values of MSE for the loss function are also shown for reference. Although there are no differences between DNN8 and DNN5, it can be seen that the values of MSE are larger when there are fewer layers, as can be seen when comparing against NN3_DO and NN3.

Table 5 shows the average of each category in order to see the tendency of the result by pattern more simply. We can easily confirm that the higher the number of layers, the higher the CORR and Direction.

## 4.2 Comparison with Support Vector Regression and Random Forests

Table 6 picks out each pattern with the highest CORR from the combination of hyper-parameters in SVR and RF, respectively, described in Section 3.5 and also picks out those pattern of neural networks from Table 4 that outperform the highest CORR of

**Table 4.** Rank correlation, directional accuracy, and mean squared error of neural networks.

| Neural Networks | CORR | Direction % | | MSE |
|---|---|---|---|---|
| | | Tertile | Quintile | |
| DNN8_1 | 0.0580 | 52.56*** | 53.36*** | **0.0834** |
| DNN8_2 | 0.0568 | 52.49*** | 53.24*** | 0.0838 |
| DNN8_3 | **0.0591** | 52.64*** | 53.37*** | **0.0834** |
| DNN8_4 | 0.0587 | **52.66*** ** | **53.48*** ** | 0.0837 |
| DNN5_1 | **0.0582** | **52.43*** ** | **53.34*** ** | **0.0833** |
| DNN5_2 | 0.0555 | 52.25*** | 53.24*** | 0.0835 |
| DNN5_3 | 0.0560 | 52.36*** | 53.22*** | 0.0835 |
| DNN5_4 | 0.0557 | **52.43*** ** | 53.26*** | 0.0836 |
| NN3_DO_1 | **0.0537** | **52.35*** ** | **52.99*** ** | **0.0839** |
| NN3_DO_2 | 0.0520 | 52.15*** | 52.75*** | 0.0840 |
| NN3_DO_3 | 0.0509 | 52.16*** | 52.94*** | 0.0841 |
| NN3_DO_4 | 0.0527 | 52.24*** | 52.87*** | 0.0841 |
| NN3_1 | 0.0450 | 52.09*** | 52.69*** | **0.0856** |
| NN3_2 | **0.0472** | 52.10*** | **53.02*** ** | **0.0856** |
| NN3_3 | 0.0437 | 51.79*** | 52.60*** | 0.0858 |
| NN3_4 | 0.0445 | **52.23*** ** | 52.61*** | 0.0859 |

***p<0.001, **p<0.01, *p<0.05.

**Table 5.** Rank correlation, directional accuracy, and mean squared error of neural networks (average).

| Neural Networks | CORR | Direction % | | MSE |
|---|---|---|---|---|
| | | Tertile | Quintile | |
| DNN8_Avg | **0.0582** | **52.59** | **53.36** | 0.0836 |
| DNN5_Avg | 0.0563 | 52.37 | 53.27 | **0.0835** |
| NN3_DO_Avg | 0.0523 | 52.23 | 52.89 | 0.0840 |
| NN3_Avg | 0.0451 | 52.05 | 52.73 | 0.0857 |

the SVR and RF patterns. The best values for each column are labeled in bold. The highest CORR from the combination of hyper-parameters in SVR is {C, gamma, epsilon} = {0.1, 0.01, 0.1}, and RF is {max_features, max_depth} = {25, 7}. For SVR and RF, we find that the tertile and quintile Direction reject the null hypothesis at the 0.1% significance level, and RF outperforms SVR including CORR. Four neural networks have been picked out, all of which are DNN, and three of which are the DNN8 patterns with the largest number of layers. In the rank relationship between CORR and Direction, DNN8_3 with the highest CORR is not completely correlated so that

**Table 6.** Rank correlation and directional accuracy of SVR, RF, and DNN.

| **Machine Learning** | **CORR** | **Direction %** | |
|---|---|---|---|
| | | Tertile | Quintile |
| SVR (best) | 0.0569 | 52.53*** | 53.30*** |
| RF (best) | 0.0576 | 52.64*** | 53.44*** |
| DNN8_1 | 0.0580 | 52.56*** | 53.36*** |
| DNN8_3 | **0.0591** | 52.64*** | 53.37*** |
| DNN8_4 | 0.0587 | **52.66*** ** | **53.48*** ** |
| DNN5_1 | 0.0582 | 52.43*** | 53.34*** |

***p<0.001, **p<0.01, *p<0.05.

DNN8_3 is not the highest Direction. It is necessary to observe carefully for this in cases where CORR does not differ much. In comparison with SVR and RF, DNN patterns outperform SVR for almost all categories, but show little superiority to RF. These results can not completely indicate the superiority of DNN, including DNN patterns which are not picked up on Table 4. However, in terms of DNN8 patterns, three patterns out of four patterns outperform in CORR and the pattern of DNN8_4 outperforms RF in all items, hence deep learning promises to be one of the leading machine learning methods.

### 4.3 Ensemble

Next, we apply ensemble methodology to combine different machine learning models and to examine whether the results improve beyond each individual pattern of Table 6. The monthly prediction scores of SVR, RF and the DNN8_3 with the highest CORR are weighted equally to create the ensemble. Table 7 shows the CORR and Direction for the tertile and quintile portfolios. We find that CORR is the highest at 0.0604, which is higher than each of the three machine learning models before combination. This demonstrates the effectiveness of the ensemble approach. For Direction, on the other hand, only quintile portfolio is the highest for the ensemble, so the improvement gained through the ensemble technique is limited.

**Table 7.** Rank correlation and directional accuracy of Ensemble.

| **Machine Learning** | **CORR** | **Direction %** | |
|---|---|---|---|
| | | Tertile | Quintile |
| Ensemble | 0.0604 | 52.56*** | 53.50*** |

***p<0.001, **p<0.01, *p<0.05.

### 4.4 Long–Short Portfolio Strategy

We have used CORR and Direction as performance measures so far, but in practice when investing based on this information, we need to analyze performance related to return more directly. We construct a portfolio strategy and use risk-adjusted return as a performance measure defined by Return/Risk (R/R) as return divided by risk. As described in Section 3.4, we construct a long–short portfolio strategy for a net-zero investment to buy top stocks and to sell bottom stocks with equal weighting in tertile and quintile portfolios. The transaction cost is not taken into account, and we examine the patterns described in Table 6 and Table 7.

The results are shown in Table 8. Return is annualized from the monthly average, and Risk is also annualized. The highest R/R is shown in bold for each tertile and quintile portfolio. We find that the highest R/R is DNN in both portfolios, DNN8_3 is 1.24 in tertile and DNN5_1 is 1.29 in quintile. Let us focus on the quintile profiles to analyze the rank relationship between Direction in Table 6, Table 7 and R/R. RF, which is higher for Direction, is the lowest for R/R, and conversely, DNN5_1 which is lower for Direction is the highest for R/R. Thus we cannot make clear conclusions.

In Table 8, some DNN patterns do not outperform SVR, RF and the ensemble so that we cannot show the complete superiority of DNN, but we can note that the pattern which has the highest R/R for each tertile and quintile comes from the DNN patterns.

## 5 Conclusions

In this paper, we implement deep learning techniques to predict one-month-ahead stock returns in the cross-section in the Japanese stock market. Our conclusions are as follows:

- In the comparison of different NN architectures, with more layers, the rank correlation coefficient (CORR) and the directional accuracy (Direction) are high. We find that DNN with greater numbers of layers could increase representational power by repeating nonlinear transformations and improve the prediction accuracy of the cross-sectional stock returns.

**Table 8.** Long–short portfolio strategy performance of SVR, RF, DNN, and Ensemble.

| Machine Learning | Tertile | | | Quintile | | |
|---|---|---|---|---|---|---|
| | Return% | Risk% | R/R | Return% | Risk% | R/R |
| SVR (best) | 8.38 | 7.46 | 1.12 | 11.36 | 9.56 | 1.19 |
| RF (best) | 9.04 | 7.37 | 1.23 | 10.72 | 9.59 | 1.12 |
| DNN8_1 | 8.88 | 7.75 | 1.15 | 11.51 | 9.69 | 1.19 |
| DNN8_3 | 9.52 | 7.70 | **1.24** | 11.59 | 9.52 | 1.22 |
| DNN8_4 | 9.33 | 7.81 | 1.19 | 12.02 | 9.93 | 1.21 |
| DNN5_1 | 8.41 | 7.56 | 1.11 | 12.32 | 9.51 | **1.29** |
| Ensemble | 8.96 | 7.45 | 1.20 | 11.59 | 9.85 | 1.18 |

- In comparison with SVR and RF, there are 4 patterns of DNN that outperform the CORR of both, while the highest Directions in each tertile and quintile are DNN patterns. Ensemble gives a limited improvement. We also examine the performance of a simple long–short portfolio strategy and find that the best R/R in each tertile and quintile portfolio is selected from DNN patterns. These results cannot completely indicate the superiority of DNN, but deep learning promises to be one of the best machine learning methods.
- We examined only 8 DNN patterns consisting of 8 layers and 5 layers compared with 24 patterns of SVR and 37 patterns of RF, and applied simple fully-connected feedforward networks. Application of recurrent neural networks, which are designed to handle time series data, is a candidate for future research. We expect that an investigation of various deep learning models could further enhance the prediction accuracy of stock returns in the cross-section.

## References

1. Subrahmanyam, A.: The cross-section of expected stock returns: What have we learnt from the past twenty-five years of research? European Financial Management 16(1), 27–42 (2010).
2. Harvey, C. R., Liu, Y., Zhu, H.: ... and the cross-section of expected returns. Review of Financial Studies 29(1), 5–68 (2016).
3. McLean, R. D., Pontiff, J.: Does academic research destroy stock return predictability? The Journal of Finance 71(1), 5–32 (2016).
4. LeCun, Y., Bengio, Y., Hinton, G.: Deep learning. Nature 521(7553), 436–444 (2015).
5. Goodfellow, I., Bengio, Y., Courville, A.: Deep Learning. MIT Press (2016).
6. Atsalakis, G. S., Valavanis, K. P.: Surveying stock market forecasting techniques–Part II: Soft computing methods. Expert Systems with Applications 36(3), 5932–5941 (2009).
7. Soni, S.: Applications of ANNs in stock market prediction: A survey. International Journal of Computer Science & Engineering Technology 2(3), 71–83 (2011).
8. Olson, D., Mossman, C.: Neural network forecasts of Canadian stock returns using accounting ratios. International Journal of Forecasting 19(3), 453–465 (2003).
9. Cao, Q., Leggio, K. B., Schniederjans, M. J.: A comparison between Fama and French's model and artificial neural networks in predicting the Chinese stock market. Computers & Operations Research 32(10), 2499–2512 (2005).
10. Kryzanowski, L., Galler, M., Wright, D.: Using artificial neural networks to pick stocks. Financial Analysts Journal 49(4), 21–27 (1993).
11. Krauss, C., Do, X. A., Huck, N.: Deep neural networks, gradient-boosted trees, random forests: Statistical arbitrage on the S&P 500. European Journal of Operational Research 259(2), 689–702 (2017).
12. Dixon, M., Klabjan, D., Bang, J. H.: Classication-based financial markets prediction using deep neural networks. CoRR(abs/1603.08604)
13. MSCI Inc. Tokyo branch.: Handbook of MSCI Index, MSCI Inc, Feb, 2017. In Japanease.
14. Kingma, D. P., Ba, J.: Adam: a method for stochastic optimization.CoRR(abs/1412.6980)
15. Smola, A. J., Schölkopf, B.: A tutorial on support vector regression. Statistics and Computing 14(3), 199–222 (2004).
16. Breiman, L.: Random forests. Machine learning 45(1), 5–32 (2001).